\documentclass[12pt]{article}
\usepackage{amsmath,latexsym,epsfig,epsf,rotate}

\newcommand{\vslash}[1]{#1\hspace{-0.5 em}/}

\begin{document}

\begin{center}
\vspace*{2cm} {\Large Vector meson dominance pion electromagnetic
form factor with the $\sigma-$model loop corrections  }\\[0pt]
\vspace{20pt} {\large S.K.~Dubinsky \\[0pt] }
\vspace{8pt}
{Kharkov National University, 61077 Kharkov, Ukraine}\\[0pt]
\vspace{10pt} {\large A.Yu.~Korchin\footnote{E-mail:
korchin@kipt.kharkov.ua},
N.P.~Merenkov\footnote{E-mail: merenkov@kipt.kharkov.ua} \\[0pt] }
\vspace{8pt}
{National Science Center ``Kharkov Institute of Physics and
Technology'',\\[0pt] 61108 Kharkov, Ukraine}
\end{center}

\begin{abstract}
A model is developed for electromagnetic form factor of the pion.
One-loop corrections are included in the linear $\sigma-$model.
The $\rho-$meson contribution is added in an extended VMD model.
The form factor, calculated without fitting parameters, is in a
good agreement with experiment for space-like and time-like photon
momenta. Loop corrections to the two-pion hadronic contribution
$a^{(had, \pi )}_{\mu}$ to the muon anomalous magnetic moment are
calculated. The optimal value of the $\sigma-$meson mass appears
to be close to the $\rho-$meson mass.
\end{abstract}

\vskip 1 cm

\section{Introduction}
\label{sec:introduction}

It has recently been understood that the pion electromagnetic (EM)
form factor is a very important physical quantity that plays a key
role in the test of the Standard Model at the electroweak
precision level. The reason is that the production cross section
\begin{equation}\label{eq:cross-section}
\sigma(e^+e^-\rightarrow
\pi^+\pi^-)=\frac{\pi\alpha^2}{3s}\bigg(1-\frac{4m_{\pi}^2}{s}\bigg)^{3/2}
|F_{\pi}(s)|^2 \ ,
\end{equation}
where $s$ is the squared total energy in center of mass system,
$\alpha$ is the fine-structure constant and $m_{\pi}$ is the pion
mass, at low energies dominates over the other hadronic channels
and accounts for more than 70 per cent of the total hadronic
contribution to the muon anomalous magnetic moment (AMM) $a_{\mu}
=(g_\mu -2)/2$. The recent measurement of $a_{\mu}$ from
Brookhaven E821 experiment \cite{Brook} has boosted the interest
in a renewed theoretical calculation of this quantity
\cite{Troconiz_01}.

The main ingredient of the theoretical prediction of $a_{\mu}$,
which is responsible for the bulk of the theoretical error, is the
hadronic contribution to the vacuum polarization. The contribution
of the  $\pi^+\pi^-$ channel to the electron--positron
annihilation process can be written in terms of the form factor
$F_{\pi}(s)$ via the dispersion integral \cite{Brodsky_68}
\begin{eqnarray}
a^{(had,\pi)}_{\mu}&=& \frac{1}{4}\int_{4m^2_\pi}^{\infty} K(s)
\bigg(1-\frac{4m_{\pi}^2}{s}\bigg)^{3/2} |F_{\pi}(s)|^2 \
\mathrm{d}s,
\label{eq:a-mu} \\
 K(s)&=&\frac{\alpha^2}{3 \pi^2 s}
\int_0^1 \frac{x^2 (1-x)}{x^2+(1-x)s/m^2_{\mu}} \mathrm{d}x \ ,
\nonumber
\end{eqnarray}
where $m_{\mu}$ is the muon mass.

 Conventional strategy of the model--independent evaluation
of this integral consists in utilization of precise experimental
data (at least at low energies, where perturbative $QCD$ cannot be
reliably applied). However, the announced accuracy, which is to be
reached soon in E821 experiment, requires calculation of EM
radiative corrections to the cross section
(\ref{eq:cross-section}) \cite{FSR}. Apart from the pure $\pi^+
\pi^-$ events, EM radiative corrections include the
$\pi^+\pi^-\gamma$ process, where photon is radiated from the
final pions. In the current experiments at $\Phi$ and $B$
factories, based on radiative return method \cite{rrm}, this
contribution cannot be extracted in a model--independent
way\footnote{Even in direct scanning experiments a
model--independent treatment of the $\pi^+\pi^-\gamma$ events
suggested in \cite{MIT} seems too complicated to be used in the
near future} and corresponding procedure requires model-dependent
approaches. This, in turn, stimulates development and study of
different theoretical models of pion-photon interaction. The
simplest one is the point--like scalar quantum electrodynamics
($sQED$) \cite{sQED} joined with standard Vector Meson Dominance
(VMD) model (see, for example, \cite{Feynman}) for description of
the $\gamma^*\rightarrow\pi^+\pi^-$ transition form factor in the
$\rho$--resonance region. Such a model was used in Ref.\cite{FSR}
for construction of the Monte Carlo event generator.

In the present paper we consider modification of the pion EM form
factor in the linear $\sigma-$model \cite{GellMann_60} with
spontaneously broken chiral symmetry, which includes the nucleon
sector. The $\rho-$meson contribution is added following Refs.
\cite{Herrmann_93,Klingle_96}. In particular, the $\rho$ coupling
to the pion and nucleon is introduced through gauge-covariant
derivatives, while direct $\gamma \rho$ coupling has explicitly
gauge-invariant form. We calculate the pion form factor in
one--loop approximation in the strong interaction and compare
$F_{\pi}(s)$ with the precise data obtained from elastic $e^-
\pi^+$ scattering and $e^+e^-$ annihilation in the pion pair.

We pay attention to the effect of the loop corrections in
$a_{\mu}^{(had,\pi)}.$ In general, as Lagrangian of the
$\sigma-$model contains Lagrangian of $sQED$ as a constituent
part, one can expect that the difference between predictions of
$\sigma-$model+VMD and $sQED$+VMD is small. Indeed, it follows
from our calculation that the loop corrections increase the
low-energy part of the right--hand--side of eq.(\ref{eq:a-mu}) by
about 2 per cent, as compared with $sQED$+VMD.

\section{\protect\bigskip Formalism}
\label{sect:formalism}
\subsection{\protect\bigskip Lagrangian}

Lagrangian of the model consists of the two parts:
${\mathcal{L}}={\mathcal{L}}^{(1)}+{\mathcal{L}}^{(2)}$. The first
one is Lagrangian of the chiral linear $\sigma-$model
\cite{GellMann_60} with an explicit symmetry-breaking term $c
\phi$. After spontaneous breaking of chiral symmetry and
re-definition of the scalar field via $\phi = \sigma + v$, where
$v = \langle \phi \rangle $ is vacuum expectation value,
Lagrangian takes the form
\begin{eqnarray}
{\mathcal{L}}^{(1)}
&=&\bar{N}(i\vslash{\partial}-m_{N})N+\displaystyle{\frac{1}{2}
\left[ {(\partial \sigma )}^{2}-m_{\sigma }^{2}{\sigma
}^{2}\right] }+\displaystyle{\frac{1}{2}\left[ {(\partial
{\vec{\pi}})}^{2}-m_{\pi
}^{2}{\vec{\pi}\,}^{2}\right] }  \notag \\
&&-g_{\pi }\bar{N}\left( \sigma +i\gamma
_{{}_{5}}\vec{\tau}\vec{\pi}\right) N-\lambda \left( {\sigma
}^{2}+{\vec{\pi}\,}^{2}\right) \left[ v\sigma
+\displaystyle{\frac{1}{4}\left( {\sigma
}^{2}+{\vec{\pi}\,}^{2}\right) }\right]  + \mathrm{const},
\label{eq:1}
\end{eqnarray}
where $N,\ \vec{\pi}\ $ and $\sigma $ are the fields of the
nucleon, pion and meson with vacuum quantum numbers, respectively,
$g_{\pi }$\ is the coupling constant, $\lambda $ is parameter of
the meson potential, and $\vslash{\partial}\equiv \partial ^{\mu
}\gamma _{\mu },$ $({\partial \sigma )}^{2}=\partial ^{\mu }\sigma
\partial _{\mu }\sigma,$ etc.
All parameters of the model are related via
\begin{equation}
m_{N}=g_{\pi }v,\quad \quad \quad m_{\sigma }^{2}=2\lambda
v^{2}+m_{\pi }^{2},\quad \quad \quad m_{\pi }^{2}=\frac{c}{v}.
\label{eq:2}
\end{equation}
Moreover, in the tree-level approximation $v=f_{\pi }$, where
$f_{\pi }=93.2$ MeV\ is the pion weak-decay constant. More details
on the $\sigma-$model can be found, for example, in
\cite{DeAlfaro}, Ch.5, sect.2.6.

The second part of Lagrangian includes interaction with the EM
field $A^{\mu },$\ and the field $\rho ^{\mu }$\ of the
\textit{neutral} $\rho-$meson. This interaction can be obtained by
using the minimal substitutions
 \begin{eqnarray}
\partial ^{\mu }N &\rightarrow & (\partial ^{\mu }+ie\frac{1+\tau
_{3}}{2}A^{\mu }+ig_{\rho }\frac{\tau _{3}}{2}\rho ^{\mu })N ,
\notag \\
\partial ^{\mu }\pi ^{a} &\rightarrow &\partial ^{\mu }\pi ^{a}+(eA^{\mu
}+g_{\rho }\rho ^{\mu })\varepsilon ^{3ab}\pi ^{b},\,\ \ \ \ \ \ \
\ \ \ \ \ \ \ \ \ \ \ \ \ (a,b=1,2,3), \label{eq:min-substitution}
\\
\partial ^{\mu } \sigma &\rightarrow & \partial ^{\mu } \sigma,
\nonumber
\end{eqnarray}
where $e$ is the proton charge, $g_{\rho }$ is the coupling
constant, and $\tau _{3}$ is the third component of the Pauli
vector $\vec{\tau}$ $=(\tau _{1},\tau _{2},\tau _{3}).$ In
addition we include the direct coupling of the photon to the $\rho
-$meson in the version of VMD model from Refs.
\cite{Herrmann_93,Klingle_96}. In this way we obtain
\begin{eqnarray} {\mathcal{L}}^{(2)}
&=&\displaystyle{\frac{1}{2}m_{\rho }^{2}\rho _{\mu }\rho ^{\mu
}-\frac{1}{4}\rho _{\mu \nu }\rho ^{\mu \nu }-\frac{1}{4}F_{\mu
\nu }F^{\mu \nu }}  \notag \\
&&-\left( eA_{\mu }+g_{\rho }\rho _{\mu }\right) {\left( \vec{\pi}\times
\partial ^{\mu }\vec{\pi}\right) }_{3}+{\left( eA_{\mu }+g_{\rho }\rho _{\mu
}\right) }^{2}({\vec{\pi}\,}^{2}-\pi _{{}_{3}}^{2})  \notag \\
&&-\displaystyle{g_{\rho }\bar{N}\gamma ^{\mu }\frac{\tau _{{}_{3}}}{2}N
\rho _{\mu }-e\bar{N}\gamma ^{\mu }\frac{1+\tau _{{}_{3}}}{2}N A_{\mu }}\
{-\frac{e}{2f_{\rho }}\,\rho _{\mu \nu }F^{\mu \nu }}.
\label{eq:rho-gamma-Lagr}
\end{eqnarray}
Here $F_{\mu \nu }=\partial _{\mu }A_{\nu }-\partial _{\nu }A_{\mu
}$, $\rho _{\mu \nu }=\partial _{\mu }\rho _{\nu }-\partial _{\nu
}\rho _{\mu }$ and $f_{\rho }$ determines the $\gamma \rho $
coupling. In eq.(\ref {eq:rho-gamma-Lagr}) we assumed equal
coupling constants of the $\rho $ to the pion and the nucleon in
accordance with the universality hypothesis of Sakurai (see, e.g.,
\cite{DeAlfaro}, Ch.5, sect.4). At the same time the $\gamma \rho
$ coupling constant $f_{\rho }$ does not necessarily coincide with
$g_{\rho }.$ Lagrangian (\ref{eq:rho-gamma-Lagr}) obeys gauge
invariance because of the form of the $\gamma \rho $ coupling. We
should mention that the nucleon contribution is also included in
Lagrangian (\ref {eq:rho-gamma-Lagr}), contrary to
\cite{Herrmann_93,Klingle_96}.

\subsection{Counterterms and renormalization}

\label{sec:counter-terms}

Since one of the purposes of the present paper is to take into
account loop corrections to the pion EM vertex one needs to
specify the way of renormalization of the parameters. We use the
conventional approach and assume that in Lagrangians (\ref{eq:1})
and (\ref{eq:rho-gamma-Lagr}) enter the ``bare'' fields, coupling
constants and masses, to be marked by the subscript ``0''. The
bare fields require rescaling: $(\vec{\pi} _{0},\sigma
_{0})=\sqrt{Z_{\pi }}(\vec{\pi} ,\sigma
),\,N_{0}=\sqrt{Z_{N}}N,\,\rho _{0}^{\mu }=\sqrt{Z_{\rho }}\rho
^{\mu }\,\ $and$\ \ A_{0}^{\mu }=\sqrt{Z_{A}}A^{\mu },$ where
$Z_{\pi ,}Z_{N},Z_{\rho }$ and $Z_{A}$ are the respective
wave-function renormalization constants for the pion (or sigma),
nucleon, rho and photon. The procedure for obtaining Lagrangian of
counterterms is known (see, for example, \cite{Peskin}, Ch.10).
For ${\mathcal{L}}^{(1)}$ the corresponding counterterm Lagrangian
reads
\begin{eqnarray}
{\mathcal{L}}_{ct}^{(1)} &=&\delta
_{Z_{N}}\bar{N}i\vslash{\partial}N-\delta _{g_{\pi
}}v\bar{N}N-\delta _{g_{\pi }}\bar{N}\left( \sigma +i\gamma
_{{}_{5}}\vec{\tau}\vec{\pi}\right) N+\frac{1}{2}\delta _{Z_{\pi
}}\left[
{(\partial {\vec{\pi}})}^{2}+{(\partial \sigma )}^{2}\right]  \notag \\
&&-\frac{1}{2}\left( \delta _{\mu }+3\delta _{\lambda }v^{2}\right) \sigma
^{2}-\frac{1}{2}\left( \delta _{\mu }+\delta _{\lambda }v^{2}\right)
{\vec{\pi}}^{2}-\frac{1}{4}\delta _{\lambda }{\left( {\vec{\pi}}^{2}+\sigma
^{2}\right) }^{2}-\delta _{\lambda }v\left( \sigma {\vec{\pi}}^{2}+\sigma
^{3}\right)  \notag \\
&&-\left[ \left( \delta _{\mu }+\delta _{\lambda }v^{2}\right) v-\delta
_{c}\right] \sigma +\mathrm{const}.  \label{eq:counter-sigma}
\end{eqnarray}
There are six constants $\ \delta _{Z_{\pi }},\delta
_{Z_{N}},\delta _{\mu },\delta _{\lambda },\delta _{g_{\pi
}},\delta _{c},$ which can be fixed by imposing, in general, six
conditions on the Green functions. In the calculation of the pion
EM vertex only one constant $\delta_{Z_\pi}$ will be needed (see
sect. \ref {sec:form-factor-sigma}).

For Lagrangian (\ref{eq:rho-gamma-Lagr}) one can define first the
physical values of the electric charge $e=e_{0}\sqrt{Z_{A}}$ and
the  rho coupling $g_{\rho }=g_{\rho 0}\sqrt{Z_{\rho }}.$ It is
also convenient to introduce $\hat{m}_{\rho }=\sqrt{Z_{\rho
}}m_{\rho 0}$ ($\rho-$meson mass in the absence of the coupling to
pions). The counterterm Lagrangian can be written as
\begin{eqnarray}
\mathcal{L}_{ct}^{(2)} &=&-\delta _{Z_{A}}{\frac{1}{4}F_{\mu \nu }F^{\mu \nu
}-}\delta _{Z_{\rho }}{\frac{1}{4}\rho _{\mu \nu }\rho ^{\mu \nu }}  \notag
\\
&&-\delta _{Z_{\pi }}\left( eA_{\mu }+g_{\rho }\rho _{\mu }\right) {\left(
\vec{\pi}\times \partial ^{\mu }\vec{\pi}\right) }_{3}+\delta _{Z_{\pi
}}{\left( eA_{\mu }+g_{\rho }\rho _{\mu }\right) }^{2}({\vec{\pi}\,}^{2}-\pi
_{{}_{3}}^{2})  \notag \\
&&-\delta _{Z_{N}}{g_{\rho }\bar{N}\gamma ^{\mu }\frac{\tau
_{{}_{3}}}{2}N\rho _{\mu }-\delta _{Z_{N}}e\bar{N}\gamma ^{\mu }\frac{1+\tau
_{{}_{3}}}{2}NA_{\mu }}\ {-\delta }_{f_{\rho }}{\frac{e}{2}\,\rho _{\mu \nu }F^{\mu \nu
}}.  \label{eq:counter-rho-gamma}
\end{eqnarray}
It is seen, that in general one needs three additional constants
$\delta _{Z_{A}},$ $\delta _{Z_{\rho }}$ and ${\delta }_{f_{\rho
}},$ once $\delta _{Z_{\pi }}$ and $\delta _{Z_{N}}$ are fixed.
Finally, the total Lagrangian is the sum\footnote{The mass
$m_{\rho }$\ in $\mathcal{L}^{(2)}$\ is replaced by $\hat{m_{\rho
}}$.}
\begin{equation}
\mathcal{L}=\mathcal{L}^{(1)}+\mathcal{L}^{(2)}+\mathcal{L}_{ct}^{(1)}+
\mathcal{L}_{ct}^{(2)}.  \label{eq:total-Lagr}
\end{equation}

\subsection{Contribution to pion EM form factor from the sector of
$\protect\sigma-$model} \label{sec:form-factor-sigma}

Feynman rules for Lagrangian (\ref{eq:total-Lagr}) are obtained
according to the standard prescriptions \cite{ItZu80}. The
counterterm constants can be found by imposing the following
constraints on the self-energy operators of the pion, sigma-meson
and nucleon respectively
\begin{eqnarray}
\Sigma _{\pi }(m_{\pi }^{2}) &=&\frac{d}{dp^{2}}\Sigma _{\pi
}(p^{2})\biggl|_{p^{2}=m_{\pi }^{2}}= \Sigma _{\sigma }(m_{\sigma
}^{2})=0,
\label{eq:costraint-self-energy} \\
\Sigma _{N}(\vslash{p})\biggl|_{\vslash{p}=m_{N}}
&=&\frac{d}{d\vslash{p}}\Sigma
_{N}(\vslash{p})\biggl|_{\vslash{p}=m_{N}}=0.  \notag
\end{eqnarray}
These conditions imply that the pole positions of the pion,
nucleon and sigma propagators are located at the physical mass of
the pion, nucleon and sigma respectively. In addition, the residue
of the pion and nucleon propagators is unity, ensuring the absence
of the renormalization for the external pion and nucleon (but not
for the external sigma-meson). We also impose the condition
$\langle \sigma \rangle =0$, which is ensured by requiring that
the so-called tadpole diagrams vanish. Correspondingly, the
tadpole diagrams will not contribute to quantities calculated
below.

In calculation of the loop integrals we used the dimensional
regularization method (see, e.g., \cite{Peskin}, Appendix A.4).
Exploiting the conditions (\ref{eq:costraint-self-energy}) we find
the constant $\delta_{Z_{\pi}}$:
\begin{equation}
\delta _{Z_{\pi }}=-\frac{g_{\pi }^{2}}{4\pi
^{2}}\int_{0}^{1}\left[ I_{\epsilon }-\ln
\frac{\tilde{\Delta}_{N\pi }}{\Lambda ^{2}}+\frac{m_{\pi
}^{2}x(1-x)}{\tilde{\Delta}_{N \pi}}+\frac{{\left( m_{\sigma
}^{2}-m_{\pi }^{2}\right)
}^{2}x(1-x)}{4m_{N}^{2}\tilde{\Delta}_{\pi \sigma }}\right]
\mathrm{d} x ,
\end{equation}
\begin{equation} \tilde{\Delta}_{N\pi }=m_{N}^{2}-m_{\pi
}^{2}x(1-x),\quad \ \ \tilde{\Delta}_{\pi \sigma }=m_{\sigma
}^{2}x+m_{\pi }^{2}(1-x)^{2}.  \label{eq:7}
\end{equation}
In these equations $I_{\epsilon }=\displaystyle{\frac{2}{\epsilon
}-\gamma _{E}+\ln 4\pi },\ \ \epsilon =4-D \rightarrow 0,\ $ where
$D$ is the space-time dimension, $\gamma _{E}\approx 0.5772 $ is
the Euler constant and $\Lambda $ is an arbitrary scale mass,
which drops out in the physical observables.

The one-loop contributions to the pion EM vertex coming from the
$\sigma-$model are shown in Fig.~1. Using the isospin structure of
the vertices, or negative charge-conjugation parity of the photon,
one can show that the diagrams ``e,f'' and ``g'' vanish. The
counterterm ``h'' cancels divergences coming from the loop
contributions ``b'' and ``c'', while the contribution ``d'' is
finite.

In general case of the off-mass-shell pions, the EM vertex $\Gamma
_{ab}^{\mu }(p_{1},p_{2},q) $ for the process $\gamma ^{\ast
}(q)\rightarrow \pi ^{a}(p_{1})+\pi ^{b}(p_{2})$ has the form
\begin{equation}
-i\Gamma _{ab}^{\mu }(p_{1},p_{2},q)=\varepsilon ^{3ab}\left[
F(p_{1}^{2},p_{2}^{2},q^{2}){(p_{2}-p_{1})}^{\mu
}+G(p_{1}^{2},p_{2}^{2},q^{2}){(p_{2}+p_{1})}^{\mu }\right] ,
\end{equation}
with scalar functions $F(p_{1}^{2},p_{2}^{2},q^{2})$ and
$G(p_{1}^{2},p_{2}^{2},q^{2}).$ On the mass shell,
$p_{1}^{2}=p_{2}^{2}=m_{\pi }^{2}$, the function $G(m_{\pi
}^{2},m_{\pi }^{2},q^{2})$ drops out, while $F(m_{\pi }^{2},m_{\pi
}^{2},q^{2})$ becomes equal to the pion form factor $F_{\pi
}(q^{2}).$ Denoting the loop corrections by \ $\Delta F^{(\sigma
)}(q^{2})$ we find
\begin{equation}
F_{\pi }^{(\sigma )}(q^{2})=1+\Delta F^{(\sigma
)}(q^{2})+\delta _{Z_{\pi }}, \label{eq:FF-pion-loops}
\end{equation}
where the total correction is finite and equal to
\begin{eqnarray}
\Delta F^{(\sigma )}(q^{2})+\delta _{Z_{\pi }}& =&\frac{g_{\pi
}^{2}}{4\pi ^{2}}\biggl[\int_{0}^{1} \ \Bigl(\ln
\frac{m_{N}^{2}-m_{\pi }^{2}x(1-x)}{m_{N}^{2}-q^{2}x(1-x)}
-x(1-x)\bigl[\frac{m_{\pi }^{2}}{\tilde{\Delta}_{N\pi
}}+\frac{{(m_{\sigma }^{2}-m_{\pi
}^{2})}^{2}}{4m_{N}^{2}\tilde{\Delta}_{\pi \sigma }}\bigr]\Bigr)
\mathrm{d}x
\nonumber \\
&& +\int_{0}^{1}\int_{0}^{1}\Bigl(\frac{y^{2}m_{\pi }^{2}}{\Delta
_{N\pi }(y,x)}+\frac{y(1-y){(m_{\sigma }^{2}-m_{\pi
}^{2})}^{2}}{4m_{N}^{2}\Delta _{\pi \sigma }(y,x)}\Bigr)
\mathrm{d}x \mathrm{d} y \biggr],
\end{eqnarray}
\begin{equation}
\begin{split}
\Delta _{N\pi }(y,x)& =m_{N}^{2}-q^{2}y^{2}x(1-x)-m_{\pi }^{2}y(1-y), \\
\Delta _{\pi \sigma }(y,x)& =m_{\sigma
}^{2}(1-y)-y^{2}(q^{2}x(1-x)-m_{\pi }^{2})
\end{split}
\end{equation}
and $\tilde{\Delta}_{N\pi },\tilde{\Delta}_{\pi \sigma }$ are
defined in eq.(\ref{eq:7}).

\begin{figure}[hbtp]
\begin{center}
\includegraphics[width=12cm]{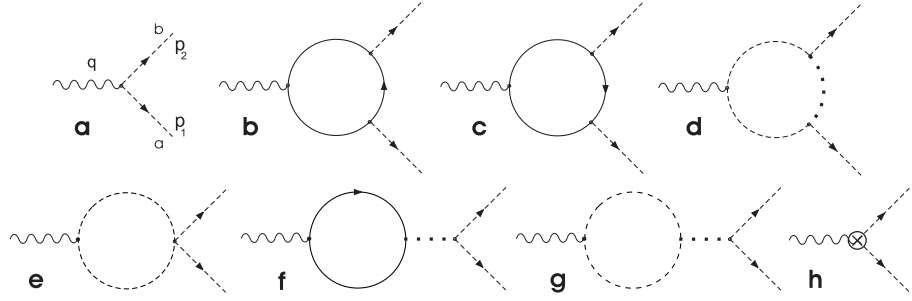}
\caption[sigma]{One-loop diagrams contributing to the pion EM form
factor in the $\protect\sigma-$model. Dashed lines depict pion,
dotted lines -- sigma, solid lines -- nucleon, and wavy lines --
photon. Small crossed circle denotes the counterterm. Diagram
``a'' corresponds to pion form factor in $sQED$. }
\end{center}
\label{Fig:1}
\end{figure}

\subsection{\protect\bigskip Contribution to pion EM form factor from the
$\protect\rho -$meson} \label{subsec:rho}

The contribution to the pion EM form factor from the $\rho-$meson
can be written in the compact form
\begin{equation} F_{\pi
}^{(\rho )}(q^{2})=-\frac{g_{\rho }(q^{2})}{f_{\rho
}(q^{2})}\frac{q^{2}}{q^{2}-\hat{m}_{\rho }^{2}-\Pi _{\rho
}(q^{2})} + \Delta F_{\pi}^{(\rho \omega) }(q^2).
\label{eq:FF-rho}
\end{equation}
This expression includes several effects coming from the loop
corrections which are shown in Fig.~2.

i) The $q^{2}-$dependent vertex $g_{\rho }(q^{2})$ describes loop
corrections to the $\rho \pi \pi $ coupling, which originate from
the $\sigma-$model (Fig.~2, diagrams ``a''). These corrections
have not been included in Ref.\cite{Klingle_96}. We can write
\begin{equation}
g_{\rho }(q^{2})=g_{\rho }[1+\Delta F^{(\sigma )}(q^{2})+\delta _{Z_{\pi }}].
\label{eq:rho-pi-pi-q2}
\end{equation}
It is seen that the expression in the square brackets is the same
as in eq.(\ref{eq:FF-pion-loops}) and is finite. From
eq.(\ref{eq:rho-pi-pi-q2}) one obtains
\begin{equation}
g_{\rho }(q^{2})=g_{\rho }(m_{\rho }^{2}) \frac{ 1+\Delta
F^{(\sigma )}(q^{2})+\delta _{Z_{\pi }}}{1+\Delta F^{(\sigma
)}(m_{\rho}^{2})+\delta _{Z_{\pi }}} \label{eq:rho-pi-pi-q2-2}
\end{equation}
in terms of the constant $g_{\rho }(m_{\rho }^{2})$. From the
experimental width of the $\rho \rightarrow \pi \pi $ decay, \
$\Gamma _{\rho \rightarrow \pi \pi }$\ = 150.7 MeV \cite{PDG}, we
get $|g_{\rho }(m_{\rho }^{2})|=6.05.$ To find the real and
imaginary part of $g_{\rho }(m_{\rho }^{2})$ one can make use of
the relations
\begin{eqnarray}
&&\mathrm{Re} \ g_\rho (m_{\rho }^2)= \frac{1}{\sqrt{1+\lambda^2}}
|g_\rho (m_{\rho }^2)|,
 \quad \quad \quad \mathrm{Im} \ g_\rho
(m_{\rho }^2)= \frac{ \lambda}{\sqrt{1+\lambda^2}} |g_\rho
(m_{\rho }^2)|,
\label{eq:Re-Im-G_rhopi} \\
&&\lambda = { \mathrm{Im} \Delta F^{(\sigma )}(m_\rho^2) } {[1+
\mathrm{Re} \Delta F^{(\sigma )}(m_\rho^2)+\delta_{Z_{\pi}
}]^{-1}}. \nonumber
\end{eqnarray}

\begin{figure}[hbtp]
\begin{center}
\epsfig{figure=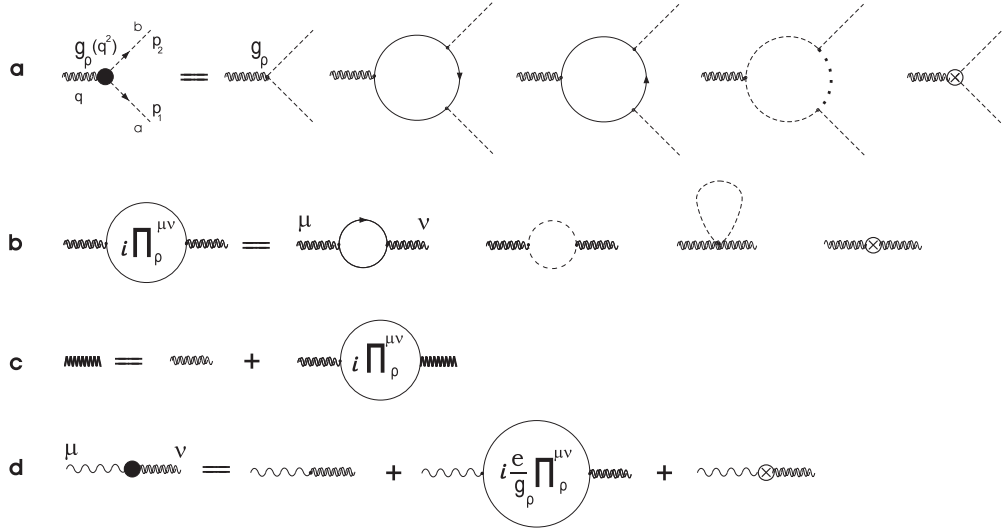,height=7cm} \caption[feyn2]{ Diagrams ``a''
show $\pi, \sigma, N$ loop corrections to $\rho \pi \pi$ vertex;
diagrams ``b'' -- $\pi, N$ loop corrections to self-energy of
$\rho-$meson; diagrams ``c'' graphically represent equation for
exact $\rho-$meson propagator; diagrams ``d'' -- $\pi, N$ loop
corrections to $\gamma \rho$ vertex. Double-wavy lines depict
$\rho-$meson. The corresponding counterterms are indicated by
crossed circles. }
\end{center}
\label{fig:2}
\end{figure}

ii) The $\rho-$meson self-energy has the structure $\ \Pi _{\rho
}^{\mu \nu }(q)=(g^{\mu \nu }-q^{\mu }q^{\nu }/q^{2})\Pi _{\rho
}(q^{2})$\ and corresponds to the diagrams shown in Fig.~2 ``b''.
\ It leads to the following exact propagator of the rho-meson
(Fig.~2, diagrams ``c'')
\begin{equation}
G_{\rho}^{\mu \nu }(q)=-i\frac{g^{\mu \nu }-q^{\mu }q^{\nu
}/q^{2}}{q^{2}-\hat{m}_{\rho }^{2}-\Pi _{\rho
}(q^{2})}+i\frac{q^{\mu }q^{\nu }}{q^{2}\hat{m}_{\rho }^{2}}.
\end{equation}
Calculation of the loop integrals in Fig.2 ``b'' results in
\begin{eqnarray}
\Pi _{\rho }(q^{2}) &=&q^{2}[\omega (q^{2})-\delta _{Z_{\rho }}],\,\ \ \ \ \
\ \ \ \ \ \ \ \ \omega (q^{2})=\frac{g_{\rho }^{2}}{16\pi ^{2}}[-I_{\epsilon
}+I(q^{2})],\,\ \   \label{eq:rho-self} \\
\ \ \ \ \ I(q^{2}) &=&2\int_{0}^{1}[2x\ln \frac{\Delta
_{N}}{\Lambda ^{2}}+(1-2x)\ln \frac{\Delta _{\pi }}{\Lambda ^{2}}]
(1-x) \mathrm{d}x, \label{eq:rho-self-energy}
\end{eqnarray}
with $\Delta _{N}=m_{N}^{2}-q^{2}x(1-x)$ and$\ \ \Delta _{\pi
}=m_{\pi }^{2}-q^{2}x(1-x).$ The self-energy has the logarithmic
divergence and needs renormalization. The authors of
Ref.\cite{Klingle_96} renormalized the self-energy by applying a
dispersion relation with two subtractions. We prefer an
alternative method of counterterms which is expressed in eq.(\ref
{eq:rho-self}) through the constant $\delta _{Z_{\rho }}.$ One can
fix the latter from the constraint on the self-energy at the
physical mass $m_{\rho } $ of the $\rho-$meson:
\begin{eqnarray}
&&\frac{d}{dq^{2}}\mathrm{Re}\ \Pi _{\rho }(q^{2})|_{q^{2}=m_{\rho
}^{2}}
=0\,\ \ \ \ \ \Rightarrow  \label{eq:derivative-zero} \\
&&\delta _{Z_{\rho }} = Z_{\rho }-1=\mathrm{Re}\ \omega (m_{\rho
}^{2})+m_{\rho }^{2}\mathrm{Re} \ \omega ^{\prime }(m_{\rho }^{2}),
\label{eq:delta-rho-counter}
\end{eqnarray}
where $\mathrm{Re}\ \omega ^{\prime }(q^{2})\equiv \frac{d}{d
q^{2}} \mathrm{Re}\ \omega (q^{2}).$ It is seen from
eqs.(\ref{eq:rho-self}) and (\ref {eq:delta-rho-counter}) that the
self-energy \ $\Pi _{\rho }(q^{2})=q^{2}[\omega
(q^{2})-\mathrm{Re}\ \omega (m_{\rho }^{2})-m_{\rho
}^{2}\mathrm{Re}\ \omega ^{\prime }(m_{\rho }^{2})]$ is finite.
Near the physical mass the latter has the expansion
\begin{equation}
\mathrm{Re} \ \Pi _{\rho }(q^{2})= -m_{\rho }^{4}\mathrm{Re}\
\omega ^{\prime }(m_{\rho }^{2}) +  {\cal O}((q^2-m_\rho^2)^2),
\label{eq:rho-self-renorm}
\end{equation}
and therefore the coupling $g_{\rho }$ is not renormalized due to
the self-energy loops \cite {Klingle_96}. There is also a finite
mass shift
\begin{equation} {m}_{\rho }^{2}-\hat{m}_{\rho }^{2} = - m_{\rho
}^{4}\mathrm{Re} \ \omega ^{\prime }(m_{\rho }^{2}).
\label{eq:mass-shift}
\end{equation}
For definition of $\hat{m}_{\rho }$ see the paragraph before
eq.(\ref{eq:counter-rho-gamma}).

Above the two-pion threshold the self-energy acquires an imaginary
part coming from the pion loop (the third diagram in Fig.~2
``b''). Namely, at $ q^{2}<4m_{N}^{2}$\ the imaginary part and the
$q^2$-dependent  $\rho \rightarrow \pi \pi $ decay width read
respectively
\begin{eqnarray} \mathrm{Im}\ \Pi (q^{2})
&=&-\frac{g_{\rho }^{2} q^2}{48\pi }(1-\frac{4m_{\pi
}^{2}}{q^{2}})^{3/2}\theta (q^{2}-4m_{\pi }^{2}),  \notag \\
\Gamma _{\rho} (q^2) &=&-\mathrm{Im}\ \Pi(q^{2}) /\sqrt{q^2}.
\label{eq:rho-decay-width}
\end{eqnarray}

iii) Closely related to the self-energy are the loop corrections
to the $\gamma \rho $ coupling constant, shown in Fig.~2 ``d''.
The sum of all contributions is proportional to the tensor \
$g^{\mu \nu }-q^{\mu }q^{\nu }/q^{2}$ similarly to the tree-level
term. Introducing the $q^{2}$-dependent vertex one obtains
\begin{equation} \frac{1}{f_{\rho }(q^{2})}=\frac{1}{f_{\rho
}}-\frac{\omega (q^{2})}{g_{\rho }}+\delta _{f_{\rho }},
\label{eq:gamma-rho-q2}
\end{equation}
where $\delta _{f_{\rho }}$ can be fixed by requiring that on the
mass shell, $q^{2}=m_{\rho }^{2}$, the coupling $f_{\rho }(m_{\rho
}^{2})$ is related to the $\rho \rightarrow e^{+}e^{-}$\ decay
width. The experimental width $\Gamma _{\rho \rightarrow
e^{+}e^{-}}=6.77$\ KeV \cite{PDG} is reproduced with $\ |f_{\rho
}(m_{\rho }^{2})|\approx 5.03$. \ From eq.(\ref{eq:gamma-rho-q2})
we find
\begin{equation}
\frac{1}{f_{\rho }(q^{2})}= \frac{1}{f_{\rho }(m_{\rho
}^{2})}+\frac{1}{g_{\rho }}[ \omega (m_{\rho }^{2})-\omega
(q^{2})]. \label{eq:gamma-rho-finite}
\end{equation}
Here the real part of the constant $f_{\rho }(m_{\rho }^{2})$ is
determined from $|f_{\rho }(m_{\rho }^{2})|$ and the imaginary
part, \ $\mathrm{Im}\ f_{\rho }(m_{\rho }^{2})$ $=  |f_{\rho
}(m_{\rho }^{2})|^2 \mathrm{Im}\ \omega (m_{\rho }^{2}) / g_{\rho
}$. It is seen from (\ref{eq:gamma-rho-finite}) that the effective
$\gamma \rho $ vertex is finite. A similar procedure for this
vertex was used in \cite {Klingle_96}, although only the real part
of $\ f_{\rho }(m_{\rho }^{2})$ was taken from experiment.

Let us also mention that in calculation of $\Pi _{\rho }(q^{2})$
and $f_{\rho }(q^{2})$ we used $|g_{\rho }(m_{\rho }^{2})|$
(instead of $g_{\rho }$) in order to get the correct width of the
$\rho-$meson.

iv) The last term in eq.(\ref{eq:FF-rho}) describes the $\rho
-\omega $ interference due to the EM effects \cite{Feynman}. The
explicit form of the contribution to the pion form factor can be
taken from Ref.\cite{OConnell_95}:
\begin{eqnarray}
\Delta F_{\pi }^{(\rho \omega )}(q^{2}) &=&-\varepsilon _{\rho \omega
}\frac{g_{\rho }}{f_{\omega }}\frac{q^{2}}{q^{2}-m_{\omega }^{2}+im_{\omega }\Gamma
_{\omega }},  \label{eq:FF-mixing} \\
\varepsilon _{\rho \omega } &=&\frac{\Pi _{\rho \omega
}}{m_{\omega }^{2}-m_{\rho }^{2}-i[ m_{\omega }\Gamma _{\omega
}-m_{\rho }\Gamma _{\rho } (q^2) ] },  \label{eq:mix-parameter}
\end{eqnarray}
where $\Gamma _{\omega }=8.43$ MeV is the full decay width of the
$\omega -$meson with mass $m_{\omega }$, \ $f_{\omega }=17.05$ is
the $\gamma \omega $ coupling constant which is fixed from the $
\omega \rightarrow e^{+}e^{-}$ decay width $\Gamma_{\omega \to e^+
e^-}=0.6$ KeV \cite{PDG}, and $ \Pi _{\rho \omega }\approx
-3.8\cdot 10^{-3}$ GeV$^{2}$ is the mixed $\rho -\omega $
self-energy.

\section{Results and discussion}
\label{sec:results}

Let us first specify parameters of the model. The constant $g_\pi$
is determined from the tree-level Goldberger-Treiman relation
$g_\pi = m_N /f_\pi$ (first equation in (\ref{eq:2})), while
$g_\rho (m_\rho^2)$ and $f_\rho (m_\rho^2)$ are fixed from
experiment, as described in sec.~\ref{subsec:rho}. The $\sigma$
mass is chosen equal to the mass of the $\rho$, i.e. $m_\sigma =
m_\rho$, in line with Ref.\cite{Weinberg_90}, where $\sigma$ and
$\rho$ are assumed to be degenerate. Furthermore, $m_{\rho}=768.5$
MeV and $m_\omega=782.57$ MeV \cite{PDG}.

It is interesting to note that calculation of the self-energy of
the $\rho-$meson gives $\hat{m}_\rho= 795$ MeV. This value is
rather close to the physical mass $m_\rho$. In this connection
note that the authors of \cite{Klingle_96} fitted $\hat{m}_{\rho
}$ from the $\pi \pi $ scattering and obtained 810 MeV. The
difference in the above values of $\hat{m}_\rho$ is partially due
to our taking into account the nucleon loop, which was not
considered in \cite{Klingle_96}.

As mentioned in sec.~\ref{subsec:rho}, both the real and imaginary
parts of the coupling constants $g_\rho (m^2_\rho)$ and $f_\rho
(m^2_\rho )$ have been included. The calculation yields:
$\mathrm{Re} \ g_{\rho }(m_{\rho }^{2}) = 6.036, \ \mathrm{Im} \
g_{\rho }(m_{\rho }^{2})= 0.405$, and $\mathrm{Re} \ f_{\rho
}(m_{\rho }^{2}) = 4.96, \ \mathrm{Im} \ f_{\rho }(m_{\rho }^{2})=
- 0.82$. Taking into account the imaginary parts leads to a small
correction to results obtained when setting $\mathrm{Im} \ g_{\rho
}(m_{\rho }^{2})= \mathrm{Im} \ f_{\rho }(m_{\rho }^{2})=0 $.

Our main results are demonstrated in Fig.~3 and Tables 1 and 2.
The calculated pion form factor $|F_{\pi}(q^2)|^2$ for space-like
and time-like values of $q^2$ is presented in Fig.~3. Apparently
the agreement with the data \cite{Amendolia_86} from elastic
electron-pion scattering, and the data \cite{Barkov_85} from
$e^-e^+$ annihilation in two pions is quite good. We emphasize
that in our approach there are no fitting or tuning parameters.

There is a strong interference of the two contributions,
$F^{(\sigma)}_\pi$ and $F^{(\rho)}_\pi$, in the total form factor.
We show in Fig.~3 separately contribution $F^{(\sigma)}_\pi$
(marked ``$\sigma-$model''). It is seen from eqs.(\ref{eq:FF-rho})
and (\ref{eq:rho-pi-pi-q2}) that the $\rho$ contribution also
includes $\pi, \sigma, N$ loops coming from the $\sigma-$model.
Switching off these corrections, i.e. putting $\Delta F^{(\sigma
)}(q^{2})+\delta _{Z_{\pi}}=0$, we obtain
\begin{equation} F_{\pi
}^{(sQED+VMD )}(q^{2})=1 -\frac{g_{\rho }}{f_{\rho
}(q^{2})}\frac{q^{2}}{q^{2}-\hat{m}_{\rho }^{2}-\Pi _{\rho
}(q^{2})} + \Delta F_{\pi}^{(\rho \omega) }(q^2).
\label{eq:FF-VMD}
\end{equation}
Here the first term corresponds to sQED, the second and the third
terms are the $\rho-$meson contributions.  Note, that
eq.(\ref{eq:FF-VMD}) corresponds to the extended version of VMD of
Ref.\cite{Klingle_96}; in the ``standard'' VMD model
\cite{Feynman} one has the dependence $\ m^2_{\rho} / [ m^2_{\rho}
- q^2 - i m_\rho \Gamma_\rho (q^2)]$.

Our calculation shows that the difference between the form factor,
calculated in $\sigma-$model+VMD (eqs.(\ref{eq:FF-pion-loops}) and
(\ref{eq:FF-rho})), and that in $sQED$+VMD (eq.(\ref{eq:FF-VMD}))
is small, and therefore the results for $sQED$+VMD are not plotted
in Fig.~3. Nevertheless, the difference may show up in the
integrated quantity $a^{(had, \pi)}_\mu$ for the muon AMM.

\begin{figure}[hbtp]
\begin{center}
\epsfig{figure=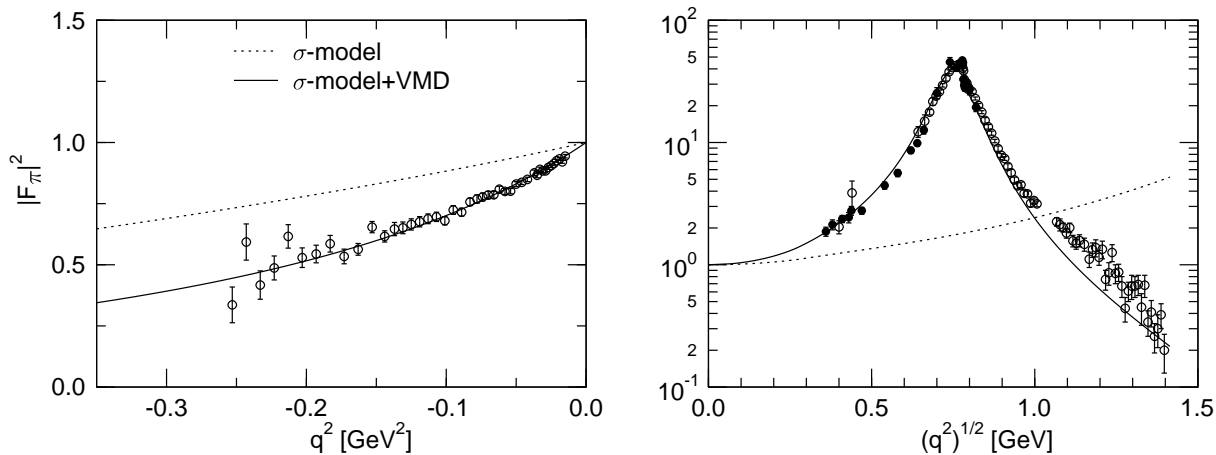,height=6cm} \caption[FF]{Pion EM form
factor for space-like $q^2$ (left) and time-like $q^2$ (right).
Experimental data are from \cite{Amendolia_86} and
\cite{Barkov_85} respectively. }
\end{center}
\label{fig:3}
\end{figure}

\begin{table}
\begin{center}
\begin{tabular}{|c|c|c|c|c|c|}
\hline $sQED$ & $\sigma-$model & $sQED$ & $\sigma-$model &
$\sigma-$model & Ref.\cite{Troconiz_01}  \\
 & &+ VMD & + VMD & + VMD ($f_\rho = g_\rho$) &  \\
\hline 525 & 753 & 4667 & 4763 & 4745 & 4774 $\pm$ 51 \\ \hline
\end{tabular}
\end{center}
\caption{Two-pion contribution $a^{(had, \pi)}_{\mu}$ to the muon
anomalous magnetic moment (in units $10^{-11}$). The upper
integration limit in eq.(\ref{eq:a-mu}) is 0.8 GeV$^2$.}
\label{tab:muon}
\end{table}

\begin{table}
\begin{center}
\begin{tabular}{|c|ccccccccc|}
\hline $m_\sigma$, GeV & 0.4 & 0.5 & 0.6 &
0.7 & 0.8 & 0.9 & 1.0 & 1.1 & 1.2  \\
\hline $a^{(had, \pi)}_{\mu}$, 10$^{-11}$ & 4546 &4583 & 4640 &
4710 & 4788 & 4867 & 4946 & 5024 & 5099
\\ \hline
\end{tabular}
\end{center}
\caption{Dependence of $a^{(had, \pi)}_{\mu}$ on the mass of
$\sigma-$meson. } \label{tab:sigma-mass}
\end{table}

The calculated values of $a^{(had, \pi)}_\mu$ are shown in
Table~\ref{tab:muon}. In general, the loop corrections in the
$\sigma-$model are important (compare the 1st and the 2nd
columns). Their role is however diminished in the full
calculation, which includes the dominant $\rho-$meson contribution
(the 3rd and the 4th columns in Table~\ref{tab:muon}). The
difference between $\sigma-$model+VMD and $sQED$+VMD calculations
is about 2\%.

In this connection our result can be used to estimate the size of
radiative corrections due to final-state-photon radiation in the
$e^+ e^- \to \pi^+ \pi^- \gamma$ process. The corresponding
contribution to the muon AMM, $a_\mu^{(had, \pi \gamma )}$, is
calculated in \cite{FSR} in framework of $sQED$+VMD. In general,
$\sigma(e^+ e^- \rightarrow \pi^+ \pi^- \gamma)/ \sigma (e^+ e^-
\rightarrow \pi^+ \pi^-) $ is of the order $ \alpha$. We expect,
that the model dependence of $a_\mu^{(had, \pi \gamma )}$ is
similar to that of $a_\mu^{(had, \pi )}$. Therefore, the deviation
of $a_\mu^{(had, \pi \gamma )}$ calculated in $sQED$+VMD, from
$a_\mu^{(had, \pi \gamma )}$ in a more realistic model, for
example, in $\sigma-$model+VMD, is about 2\%. The overall
model-dependence effect in the contribution $a_\mu^{(had, \pi
\gamma )}$ to the muon AMM is of the order $ \alpha \times 2\%
\approx 0.015\%$ and is therefore negligible.

In the 5th column we show result obtained if we put $f_\rho =
g_\rho$ in Lagrangian (\ref{eq:rho-gamma-Lagr}). This
approximation corresponds to the full universality of Sakurai. In
this case the renormalization procedure for the $\gamma \rho$
vertex changes and $\delta_{f_\rho}$ in
eq.(\ref{eq:counter-rho-gamma}) is equal to $\delta_{Z_\rho}/
g_\rho$. The numerical results for the form factor and $a^{(had,
\pi)}_\mu$, however, change very little, e.g., for the integral by
about 0.4\%.

As mentioned above, in the calculation we chose the mass
$m_\sigma=m_\rho$ for the $\sigma-$meson. This particle is
associated with $f_0 (400-1200)-$meson in \cite{PDG}. In view of
its undetermined status we study in Table~\ref{tab:sigma-mass} the
dependence of the calculated integral $a^{(had, \pi)}_{\mu}$ on
$m_\sigma$. As can be seen from Table~\ref{tab:sigma-mass}, the
integral varies considerably. Let us take the value 4774 $\pm$ 51
from \cite{Troconiz_01} as a very accurate fit to the experimental
integral. Then, for the indicated error bars we obtain the mass of
the $\sigma$ in the interval from 720 to 850 MeV. The central
value 785 MeV is surprisingly close to the $\rho-$meson mass.
Therefore our calculation agrees with the hypothesis of
Ref.\cite{Weinberg_90} about degeneracy of $\sigma-$ and
$\rho-$mesons.

In the calculations we took into account diagrams, where the
$\rho-$meson enters only on the tree level. In particular, the
loops with intermediate $\rho-$meson for the $\gamma \pi^+ \pi^-$
and $\rho \pi^+ \pi^-$ vertices are left out. Such contributions
can be consistently considered in the models, in which
$\rho-$meson is included together with its chiral partner, the
axial-vector $a_1-$meson, for example, in the so-called gauged
$\sigma-$model \cite{Lee_68} or chiral quantum hadrodynamics
\cite{Serot_92,Korchin_03}. This work is planned for future.

\section{Conclusions}
\label{sec:conclusions}

We developed a model for EM vertex of the pion. The model is based
on the linear $\sigma-$model, which generates the loops with
intermediate pion, sigma and nucleon. The $\rho-$meson is included
in line with an extended VMD model \cite{Klingle_96}. The coupling
of the $\rho$ to the pion and nucleon is introduced through
gauge-covariant derivatives, and the direct $\gamma \rho$ coupling
has gauge-invariant form. The $\rho-$meson self-energy and
modified $\gamma \rho$ vertex are generated by the pion and
nucleon loops. The renormalization is consistently performed using
the method of counterterms without cut-off parameters.

The pion EM form factor, calculated in one-loop approximation in
the strong interaction, is in good agreement with the precise data
obtained from elastic $e^- \pi^+$ scattering and $e^+e^-$
annihilation into $\pi^+ \pi^-$. The effect of the $\sigma-$model
loops turns out to be small.

We calculated the contribution of the $e^+e^- \rightarrow \pi^+
\pi^-$ process to the muon AMM moment, $a^{(had, \pi)}_{\mu}$. The
calculation agrees quite well (by 0.15\%) with recent very
accurate fit of Ref.\cite{Troconiz_01}. The contribution of the
$\sigma-$model loops to $a^{(had, \pi)}_{\mu}$ is about 2\%.

We also estimated the size of the model-dependence effects in
$a^{(had, \pi \gamma)}_{\mu}$, the contribution  to the muon AMM
from final-state-photon radiation in $e^+e^- \rightarrow \pi^+
\pi^- \gamma$ process. It is about $\alpha \times 2\% \approx
0.015\%$ and is therefore negligible. Thus our calculation does
not contradict to the conclusion of Ref.\cite{FSR} that
final-state radiative process $e^+e^- \rightarrow \pi^+ \pi^-
\gamma$ can be evaluated in scalar QED supplemented with VMD
model.

The only free parameter of the model, which is not fixed from the
experiment, is the $\sigma-$meson mass. Comparison with the fit of
Ref.\cite{Troconiz_01} strongly points to the value of this mass
close to the mass of the $\rho-$meson. This conclusion is
consistent with Ref.\cite{Weinberg_90}, where the mesons $\sigma$
and $\rho$ are assumed to be degenerate.


\end{document}